\documentclass[twocolumn,aps,prl,showpacs]{revtex4-1}
\usepackage[latin9]{inputenc}
\usepackage{amsmath}
\usepackage{amssymb}
\usepackage{graphicx}

\begin{document}

\title{Stability and Anomalous Compressibility of Bose Gases Near Resonance: \\ The scale-dependent interactions and thermal effects}

\author{Shao-Jian Jiang}
\affiliation{Department of Physics and Astronomy, University of British
	Columbia, Vancouver V6T 1Z1, Canada}
\author{Fei Zhou}
\affiliation{Department of Physics and Astronomy, University of British
	Columbia, Vancouver V6T 1Z1, Canada}

\date{Mar 10, 2015}
\begin{abstract}
The stability of Bose gases near resonance has been a puzzling problem in recent years.
In this Letter, we demonstrate that in addition to generating thermal pressure, thermal atoms enhance the repulsiveness of the scale-dependent interactions between condensed atoms due to renormalization effect and further stabilize the Bose gases.
Consequently, we find that, as a precursor of instability, the compressibility develops an anomalous structure as a function of scattering length and is drastically reduced compared with the mean-field value.
Furthermore, the density profile of a Bose gas in a harmonic trap is found to develop a flat top near the center. This is due to the anomalous behavior of compressibility and can be a potential smoking gun for probing such an effect.
\end{abstract}


\maketitle

Recently, interest in interacting quantum gases has been revived due to the application of Feshbach resonance \cite{inouye1998,cornish2000,Strecker2002a,Khaykovich2002a,chin2010}. 
Feshbach resonance provides the tunability of the scattering length $a$, which uniquely characterizes the low-energy inter-particle interaction, from zero to infinity and also from negative to positive.
It offers an easy experimental access to resonant quantum gases.
Since the conventional perturbation theories are no longer valid for these strongly interacting systems, we are confronted with the theoretical challenge of unravelling the puzzle of large-scattering-length physics. 
Among the various applications of Feshbach resonance, what has attracted particular interest is the atomic Bose gas at large positive scattering lengths, known as the resonant Bose gas on the upper branch of a Feshbach resonance \cite{papp2008,pollack2009,navon2011,wild2012,ha2013,fletcher2013,makotyn2014,yin2013,Kain2014a,rancon2014,Sykes2014,liu2015,rem2013}. 

For a dilute Bose gas where the scattering length is small and positive, the dominant contribution to its chemical potential is described by the Hartree-Fock mean-field value, $\mu_{\text{HF}} = 4 \pi a n$, with $n$ the number density of the particles.
The leading order correction to the mean-field result 
is proportional to a dimensionless parameter $\sqrt{na^3}$, and the dilute limit is hence defined as $na^3 \ll 1$. 
Near resonance where $n a^3 \sim 1$ or $\gg 1$, the dilute-gas theory \cite{lee1957,lee1957b,Beliaev58,wu1959,sawada1959,Pines59,braaten2002,Fetter1971a,abrikosov1975methods} is not applicable and  several nonperturbative approaches have been taken to understand resonant many-body physics.
They all predict the fermionization of bosons near resonance, which denotes that the chemical potential of the Bose gas reaches nearly the Fermi energy of a Fermi gas with the same density \cite{cowell2002,song2009,diederix2011}. 
In other words, the chemical potential of a Bose gas is predicted to increase linearly with $a$ in the mean-field regime when $n a^3 \ll 1$ and saturate at an energy scale defined by $n$ as resonance is approached.
These theory predictions so far are consistent with experimental studies on chemical potentials \cite{navon2011}.

One of the major challenges of creating and probing unitary Bose gases experimentally comes from the short lifetime of the gases. 
Overcoming this difficulty has been one of the main focuses of a few  recent experimental efforts \cite{fletcher2013, rem2013, makotyn2014}. 
Especially, a very interesting attempt was made a year ago by quenching the scattering length and stabilizing the gas for relatively long time via a dynamic approach \cite{makotyn2014}.
It had been generally believed that the short lifetime can be solely attributed to the few-body losses, which tend to increase rapidly when a resonance is approached.
On the other hand, recent theoretical works suggest a more dramatic onset of many-body instability beyond a critical scattering length at zero temperature \cite{borzov2012, zhou2013}.
This quantum critical point is mainly dictated by the scale dependence of interactions or the running of the coupling constant. 
Beyond the critical point, the coupling constant changes its sign at a relevant many-body energy scale, resulting in an effectively attractive interaction between condensed atoms and hence an instability.
The competition between few-body losses and the onset of many-body instability remains to be further studied in future experiments.

Given the limited current understanding of this issue, a few questions that are both fundamentally and practically interesting arise.
1) What is the stability domain for Bose gases in the $T-a$ plane? 2) What are the mechanisms for stabilization at finite temperatures?
3) What anomalous properties does a Bose gas have near the stability domain?
4) What can be used as a smoking gun for the stability boundary in experiments?
In this Letter, we present an investigation of this topic and address the four questions raised above. 
Although there have been several previous theoretical studies
on near-resonance Bose gases, they were exclusively focused on the zero temperature case \cite{cowell2002,pilati2005,diederix2011,song2009,borzov2012,Mashayekhi2013,zhou2013}. In contrast, the
finite but low temperature case, which is more closely
related to experiments, demands further investigation.

\begin{figure}[t]
	\includegraphics[width=3.5in]{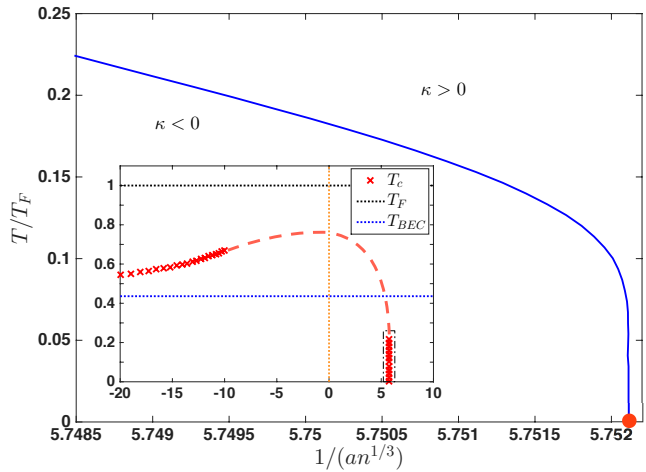}
	\caption{(color online). The phase diagram of a scattering atomic Bose gas near the $T=0$ quantum critical point. 
Temperature is shown in unit of the ``Fermi temperature'' $T_F$ defined for a Bose gas with density $n$, which is equal to $(6 \pi^2)^{2/3}/2 n^{2/3} \approx 7.6 n^{2/3}$.
The blue solid line is where the compressibility $\kappa$ changes its sign. 
It separates the stable and unstable regions where the compressibility is positive and negative respectively.
The red dot on the horizontal axis represents the quantum critical point.
The inset shows the overall phase diagram in the same parameter space with both positive and negative scattering lengths, where the red crosses represent the data from our calculation.
The red dashed line connecting the data is drawn as a guide for the eyes in the vicinity of resonance.
The dot-dashed box marks the corresponding region in the main plot. 
For reference, we plot in blue and black dotted lines the BEC transition temperature for an ideal Bose gas $T_{BEC}$ and $T_F$ respectively. 
The yellow dotted line shows the position of resonance.}
	\label{phase}
\end{figure}

Fig.~\ref{phase} shows the phase diagram for resonant Bose gases.
The main plot displays the phase diagram near the quantum critical point (the red dot on the horizontal axis). 
The phase boundary is defined as the line where the compressibility changes its sign.
It separates the regions with positive and negative compressibility.
As shown in Fig.~\ref{phase}, the critical scattering length increases as the temperature rises from zero, indicating that small finite temperatures stabilize the gas.
In our calculations, the compressibility $\kappa$ is calculated via,
\begin{equation}
\label{eq:7}
\frac{1}{\kappa} = \frac{\partial \mu}{\partial n}
\end{equation}
from the equation of state $\mu(n)$.
$\mu$ is obtained from $\mu = \partial F(n_0, \mu)/ \partial n_0$, where $n_0$ is the number density of condensed atoms and $F(n_0, \mu)$ is the free-energy density \cite{Beliaev58, Pines59}. 
$F(n_0, \mu)$ itself is calculated at a prefixed chemical potential, thus $\mu$ is determined self-consistently (refer to Eq.~\eqref{eq:3} and the context there).
In the low-temperature limit, 
we have identified two dominant processes 
that yield the leading order contributions to $F$.
We denote them as $F \approx F_1 + F_2$. $F_1$ has the following form,
\begin{equation}
\label{eq:2}
F_1 = T \int \frac{d^3 q}{(2 \pi)^3} \ln( 1 - e^{-\beta (\epsilon_k + \eta)}),
\end{equation}
where $\epsilon_k = k^2/2$ is the energy of a free particle, $\eta = \Sigma - \mu$, $\Sigma$ is the self energy, and $\beta = 1/T$.
For clarity, we set the reduced Planck constant $\hbar$, the Boltzmann constant $k_B$, and the atomic mass $m$ to be unity throughout the paper.
$F_1$ counts the contributions of scattering thermal atoms whose energy receives a modification of $\eta$ due to many-body interactions.
As expected, the thermal pressure provided by $F_1$ enhances the thermodynamical stability of the Bose gas.
$F_2$ on the other hand can be attributed to the interaction energy of the condensed atoms and can be written as 
\begin{eqnarray}
\label{eq:1}
F_2 & = & \frac{1}{2} g_2(\eta) n_0^2,
\end{eqnarray}
where $g_2(\eta)$ corresponds to the renormalized two-body running coupling constant at the energy scale defined by $\eta$ \cite{zhou2013}.
At finite temperatures, $g_2(\eta)$ has the form
\begin{equation}
\label{eq:8}
g_2^{-1}(\eta) = \frac{1}{4 \pi a} - \frac{\sqrt{2 \eta}}{4 \pi} + \int \frac{d^3 q}{(2 \pi)^3} \frac{2 n_B(q^2/2 + \eta)}{q^2 + 2 \eta},
\end{equation}
where $n_B(x) = (e^{\beta x} - 1)^{-1}$ is the bosonic distribution function. 
The last term of Eq.~\eqref{eq:8} represents the bosonic-enhancement effect due to thermally excited atoms.
This makes the two-body interaction more repulsive and in turn stabilizes the gas.
In the limit of low temperature, we also perform an analytical expansion of Eqs.~\eqref{eq:2} and \eqref{eq:1}.
Our analysis shows that 
the thermal pressure is the driving force that stabilizes the gas only when $T$ is small.
At $T \approx 0.17T_F$, the contribution of the running coupling constant becomes comparable to that of thermal pressure and hence more dominant.

When the scattering length is tuned to be negative, the interaction between atoms is effectively attractive. 
At low temperatures, the gas is thermodynamically unstable. 
It has been shown that for a small negative scattering length, Bose gases become stable at a temperature slightly above the transition temperature of Bose-Einstein condensate (BEC) for an ideal gas, $T_{BEC}$ \cite{stoof1994,mueller2000,jeon2002}.
This result can also be obtained using our approach outlined here.
When $T>T_{BEC}$, $n$ is related to $F$ by $n= - \partial F / \partial \mu$.
By further supplementing an equation for the self energy $\Sigma = 8 \pi a n$, we can identify the phase boundary for small negative scattering lengths.
Now we  present a more thorough phase diagram for Bose gases, shown in the inset of Fig.~\ref{phase}. 
The red crosses mark the instability points where $\kappa$ changes its sign for both small negative scattering lengths and also around the quantum critical point.
Near resonance, we extrapolate between these two cases by smoothly connecting the two sides.
The resultant curve (red dashed line) near $a=\infty$ is qualitatively consistent with the picture in Ref.~\cite{li2012} derived via a high-temperature expansion.
\begin{figure}[t]
	\includegraphics[width=3.6in]{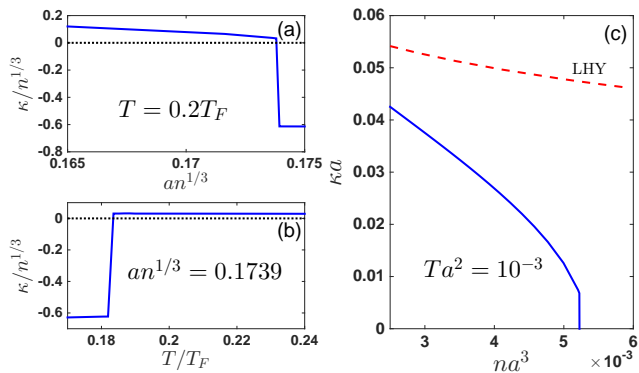}
	\caption{(color online). The change of sign of compressibility as functions of (a) scattering length when temperature $T= 0.2 T_F$ and (b) temperature when scattering length $a=0.1739 n^{-1/3}$. (c) Compressibility as a function of density for fixed scatteirng length at temperature $T=10^{-3} a^{-2}$ (blue solid line). The compressibility drops rapidly near the critical $na^3$, compared with the result from Lee-Huang-Yang (LHY) theory \cite{lee1957,lee1957b} (red dashed line).
}
	\label{cmprs}
\end{figure}

\begin{figure}[t]
	\includegraphics[width=3.5in]{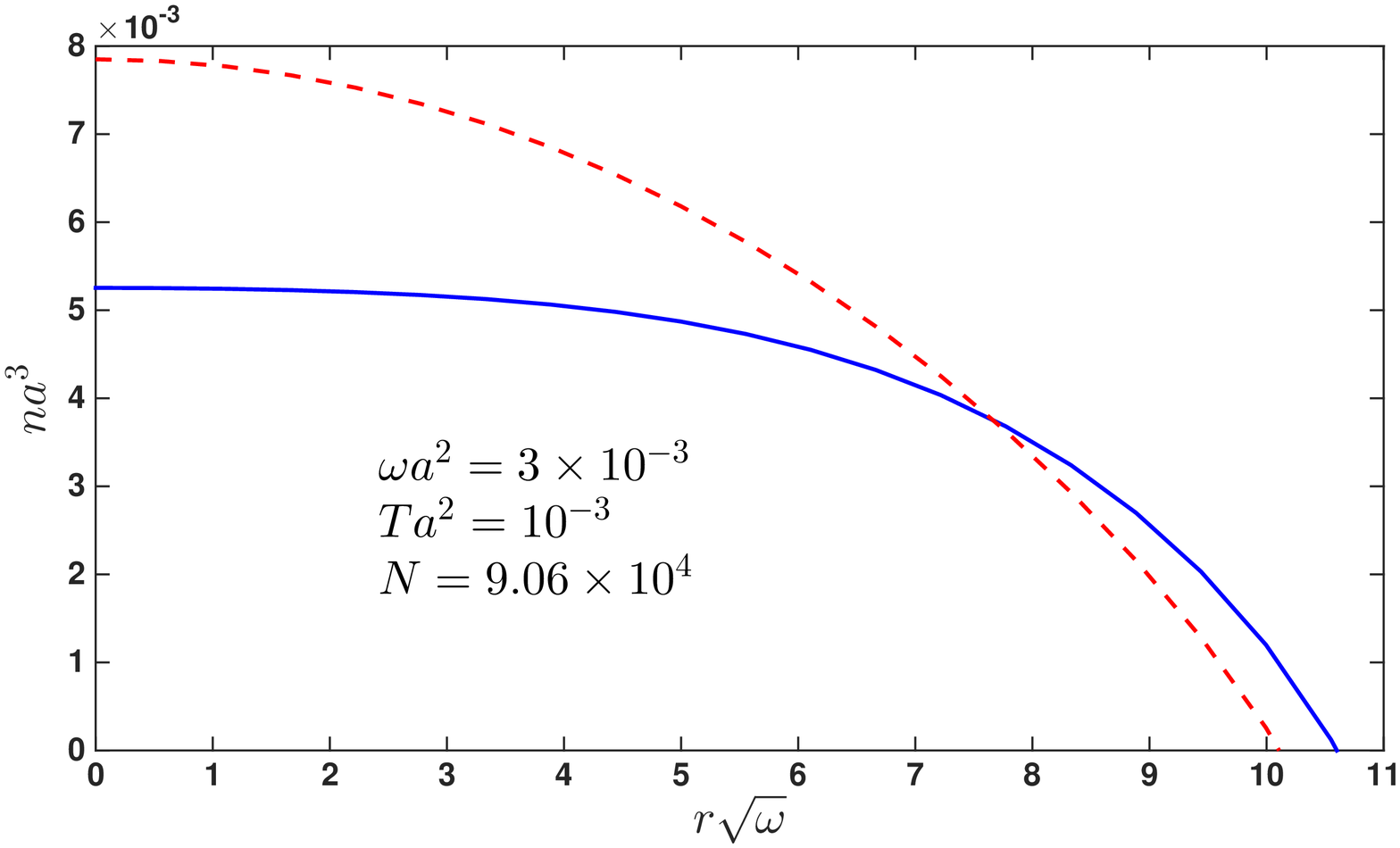}
	\caption{(color online). Density profile for $N = 9.06\times 10^4$ atoms in a harmonic trap with frequency $\omega = 3 \times 10^{-3} a^{-2}$.
The blue solid line shows the result of our calculation within local-density approximation at temperature $T = 10^{-3} a^{-2}$.
The density in the trap center is $na^3 = 5.25 \times 10^{-3}$, slightly below the critical value.
Note that the rapid drop of compressibility near $na^3 = 5.25 \times 10^{-3}$ (see Fig.~\ref{cmprs} (c)) leads to the flat top of the density profile.
We also plot the density profile from LHY theory in red dashed line for comparison.
}
	\label{density}
\end{figure}

Now we turn to questions 3) and 4) asked in the introduction regarding the experimental signature of the stability boundary.
A previous study on two dimensional Bose gases \cite{Mashayekhi2013} which applied a theoretical approach similar to the one outlined here implies an anomalous behavior of the compressibility.
That is the inverse of compressibility as a function of scattering length develops a maximum as a precursor of the instability.
This is consistent with a recent experiment in Chicago on two dimensional Bose gases \cite{ha2013} where a maximum in the inverse of compressibility was measured.
Given the success in two dimension, here we further show in Fig.~\ref{cmprs} the data for the anomalous compressibility along the phase boundary. 
Figs.~\ref{cmprs} (a) and (b) show the sign change of the compressibility. Before reaching the instability point, the system has a positive compressibility and is thermodynamically stable.
As a precursor of the instability, the compressibility decreases rapidly as shown in Fig.~\ref{cmprs} (c).

Furthermore, since the gas with a negative compressibility is difficult to control, it is appealing to ask what is the potential smoking gun for the stability boundary in a three dimensional Bose gas before it becomes unstable.
For trapped Bose gases, this anomalous compressibility leads to a distinct feature in the density profile.
Within the local-density approximation the trap renders a position-dependent chemical potential, which reaches its maximum value in the center of the trap and decreases towards the edge \cite{Dalfovo1999a}. 
Consequently, the density profile of the gas, $n(\boldsymbol{r})$, is a direct reflection of the compressibility and can provide experimental evidence of the quantum critical behavior discussed above. 
In Fig.~\ref{density}, we plot the density profile for a Bose gas in a harmonic trap.
The peak density, which is located in the center of the trap, is set slightly below the critical value $n a^3 = 5.25 \times 10^{-3}$. 
$N=9.06 \times 10^4$ is the critical particle number for this set of $\omega$ (the trap frequency), $T$, and $a$, beyond which  the gas near the trap center becomes unstable.
In general, we find that $N_{\text{cr}} \approx (\sqrt{\omega}a)^{-6} f(Ta^2)$, where $N_{\text{cr}}$ is the critical particle number and $f$ is a dimensionless function and $f(0) =  0.0024$.
For comparison, we also plot in the red dashed line the density profile obtained using the Lee-Huang-Yang (LHY) theory \cite{lee1957,lee1957b} for the same $\omega$, $a$, and $N$ at zero temperature.
It can be seen that our calculation predicts a much flatter top for the density profile compared with the LHY result.
This is due to the rapid drop of $\kappa$ near the critical point (i.e., the gas becomes difficult to compress before reaching the instability), in contrast to the mild decrease in the LHY theory, as shown in Fig.~\ref{cmprs} (c).
There have been experimental evidences of a flat top of the density profile for a Bose gas in a harmonic trap near resonance, but they have long been attributed to the strong three-body loss in the center of the trap. 
However, our analysis shows that the worrying three-body loss is still relatively small near the instability point, which we will discuss in more details below.
As a result, the observed flat top is a strong evidence of the anomalous compressibility, a precursor of the onset of many-body instability.

Bose gases of scattering atoms suffer three-body-recombination processes due to the presence of bound states deeper in the energy spectrum.
When three atoms scatter, two atoms can form a bound state while the third atom gains appreciable kinetic energy and may escape from the trap; this is known as three-body loss.
The gas therefore has a finite lifetime.
The lifetime due to losses is determined by the rate of this three-body-recombination event, known as the three-body loss rate $L_3$.
There have been extensive studies on $L_3$ \cite{esry1999,bedaque2000,fletcher2013,fedichev1996,nielsen1999,rem2013}.
It was shown that,  $L_3$ increases as $a^4$ (apart from the peaks and dips due to interference effects) and saturates to $\sim T^{-2}$ when $a$ becomes comparable to the thermal wave length.
This smooth variation of $L_3$ with $a$ is fundamentally different from the sharp sign change of $\kappa$ at the onset of the many-body instability discussed above.
In order to further compare the few- and many-body effects, we carry out an estimation of $L_3$ near the quantum critical point.
In the low-temperature limit, $L_3$ was shown to be $203.7 a^4$ up to a periodic function of $\ln (a \Lambda)$, oscillating between $0$ and $1$, where $\Lambda$ is an ultraviolet momentum scale depending on the short-distance details of the interatomic interaction \cite{bedaque2000}.
Near the critical point where $a n^{1/3} \approx 0.174$, the energyscale set by the three-body loss rate can be estimated as $L_3 n^2 \approx  0.0246 T_F$, while the chemical potential near this point is calculated to be $\mu \approx 0.735 T_F$, i.e., the onset of many-body instability sets in before the three-body loss becomes significant.
This is in contrast to the common belief that few-body loss is the main limitation to the experimental accessibility of highly degenerate Bose gases.
On the other hand, we can also infer that, compared with the many-body timescale, the lifetime due to three-body loss is long near the critical point.
This validates our treatment of the system as a thermodynamical system and justifies our thermodynamical analysis above.
Besides, the presence of the ultraviolet parameter $\Lambda$ in $L_3$ reflects the nonuniversality of Bose gases near resonance, which is related to the Efimov physics \cite{Efimov70}.
This effect can be incorporated in the effective-field-theory approach quite straightforwardly at zero temperature by including the three-body interaction energy with an ultraviolet cutoff \cite{borzov2012}.
By varying the ultraviolet cutoff from $10^{-2} n^{1/3}$ to $10^{-3} n^{1/3}$, our calculation shows an oscillation of the position of the critical point, indicating nonuniversality, but the relative magnitude of the oscillation is less than $2\%$.
This suggests that the three-body effect is negligible even near the critical point, which is consistent with previous studies. 

\label{sec:method}
At last, we briefly introduce our technique. We adopt an effective-field-theory approach combined with self-consistent equations \cite{borzov2012,zhou2013}.
In this approach, the free-energy density $F(n_0, \mu)$ (which reduces to the energy density at zero temperature) is calculated as a function of condensed-particle density $n_0$ and a preassumed chemical potential $\mu$.
At equilibrium, $\mu$ is further related to $F(n_0,\mu)$ by
\begin{equation}
\label{eq:3}
\mu = \frac{\partial F(n_0, \mu)}{ \partial n_0}, n = n_0 - \frac{\partial F(n_0,\mu)}{\partial \mu},
\end{equation}
where the second equation is the number equation.
The chemical potential, condensation fraction, and thermodynamical quantities like compressibility can be obtained by solving Eq.~\eqref{eq:3} self-consistently. 

The evaluation of $F$ can be carried out diagrammatically \cite{abrikosov1975methods,Fetter1971a,Beliaev58,Pines59,mahan2000many}:
\begin{equation}
\label{eq:5}
F(n_0, \mu) = \sum_{M=0}^{\infty}\frac{g_M^{(0)}(\mu)}{M!}n_0^M,
\end{equation}
where $g_M^{(0)}(\mu)$ is the effective $M$-body interaction.
However, this traditional perturbation expansion converges quite slowly.
In our approach, we further introduce the irreducible $M$-body interactions which depend on $n_0$ themselves \cite{borzov2012}.
In this case, $F$ can be rewritten as,
\begin{equation}
\label{eq:6}
F(n_0, \mu) = \sum_{M=0}^{\infty}\frac{g_M^{(\text{IR})}(\mu,n_0)}{M!}n_0^M,
\end{equation}
where $g_M^{(\text{IR})}$ is the irreducible $M$-body interaction.
At zero temperature, this summation starts with $M=2$, and $g_2^{(\text{IR})}$ has been shown to produce more than $99\%$ of the LHY correction in the dilute limit.
Near resonance, this expansion has also been proven to converge rapidly \cite{borzov2012}.
According to our calculations at zero temperature, the contribution of $g_3^{(\text{IR})}$ to $F$ near the critical point is less than $5\%$ of that of $g_2^{(\text{IR})}$.
A rigorous solution at $4-\epsilon$ dimensions \cite{jiang2014} further demonstrated that the effect of $M$-body ($M \ge 3$) interactions is indeed suppressed by an extra power of $\epsilon$, which, after extrapolation, is consistent with the numerical smallness we observed here.
In our present analysis, we neglect contributions with $M \ge 3$.
On the other hand, there are contributions from $g_0^{(\text{IR})}$ at finite temperatures, which correspond to the free energy of thermal atoms.
Among this type of diagrams, those involving a single thermal atom give the leading-order contribution  while others are suppressed by additional $e^{-\beta \mu}$ factors.
For example, the Nozieres-Schmitt-Rink type diagrams considered in Ref.~\cite{li2012} are suppressed in the low-temperature limit.

In summary, we have studied the compressibility and stability of the resonant atomic Bose gas.
Near the instability line, besides generating a thermal pressure, thermal excitations further stabilize the gas by enhancing the repulsiveness of the interatomic interactions.
Unlike the smooth varying of lifetime due to few-body loss, this onset of instability due to the many-body effects is characterized by a sharp sign change of the compressibility.
As a precursor of the instability, the compressibility drops quickly, which induces a flat top in the density profile in the presence of a harmonic trap.
This might be a potential smoking gun for the onset of many-body instability.
It is worth emphasizing that the instability studied here occurs when the few-body losses are still insignificant; this makes it plausible to experimentally detect this many-body instability.
Our study provides useful insight on the thermodynamics of scattering atomic Bose gases for a wide range of scattering lengths and can further shed light on future research on the dynamics of Bose gases.

\begin{acknowledgments}
This work is supported by the Canadian Institute for Advanced Research under the Quantum Materials Program. We thank Christopher Chamberland for his early contributions to this work, Jeff Maki for proofreading the manuscript, and Jun-Liang Song for sharing his codes. One of the authors (F. Z.) thanks C. Chin, E. Cornell, R. Hulet, D. Jin, and C. Salomon for inspiring discussions about the flat top physics.
\end{acknowledgments}


\begin{thebibliography}{References} 

\bibitem{inouye1998} S. Inouye, M. R. Andrews, J. Stenger, H.-J. Miesner, D. M. Stamper-Kurn, and W. Ketterle, Nature {\bf 392}, 151 (1998).

\bibitem{cornish2000} S. L. Cornish, N. R. Claussen, J. L. Roberts, E. A. Cornell, and C. E. Wieman, Phys. Rev. Lett. {\bf 85}, 1795 (2000).

\bibitem{Strecker2002a} K. E. Strecker, G. B. Partridge, A. G. Truscott, and R. G. Hulet, Nature {\bf 417}, 150 (2002).

\bibitem{Khaykovich2002a} L. Khaykovich, F. Schreck, G. Ferrari, T. Bourdel, J. Cubizolles, L. D. Carr, Y. Castin, and C. Salomon, Science {\bf 296}, 1290 (2002).

\bibitem{chin2010} C. Chin, R. Grimm, P. Julienne, and E. Tiesinga, Rev. Mod. Phys. {\bf 82}, 1225 (2010).







\bibitem{papp2008} S. B. Papp, J. M. Pino, R. J. Wild, S. Ronen, C. E. Wieman, D. S. Jin, and E. A. Cornell, Phys. Rev. Lett. {\bf 101}, 135301 (2008).

\bibitem{pollack2009} S. E. Pollack, D. Dries, M. Junker, Y. P. Chen, T. A. Corcovilos, and R. G. Hulet, Phys. Rev. Lett. {\bf 102}, 090402 (2009).

\bibitem{navon2011} N. Navon, S. Piatecki, K. G\"{u}nter, B. Rem, T. C. Nguyen, F. Chevy, W. Krauth, and C. Salomon, Phys. Rev. Lett. {\bf 107}, 135301 (2011).

\bibitem{wild2012} R. J. Wild, P. Makotyn, J. M. Pino, E. A. Cornell, and D. S. Jin, Phys. Rev. Lett. {\bf 108}, 145305 (2012).

\bibitem{ha2013} L.-C. Ha, C.-L. Hung, X. Zhang, U. Eismann, S.-K. Tung, and C. Chin, Phys. Rev. Lett. {\bf 110}, 145302 (2013).

\bibitem{fletcher2013} R.J. Fletcher, A.L. Gaunt, N. Navon, R.P. Smith, and Z. Hadzibabic, Phys. Rev. Lett. {\bf 111}, 125303 (2013).

\bibitem{yin2013} X. Yin and L. Radzihovsky, Phys. Rev. A {\bf 88}, 063611 (2013).
\bibitem{Kain2014a} B. Kain and H. Y. Ling, Phys. Rev. A {\bf 90}, 063626 (2014).


\bibitem{makotyn2014} P. Makotyn, C. E. Klauss, D. L. Goldberger, E. A. Cornell, and D. S. Jin, Nat. Phys. {\bf 10}, 116 (2014).
\bibitem{Sykes2014} A. G. Sykes, J. P. Corson, J. P. D'Incao, A. P. Koller, C. H. Greene, A. M. Rey, K. R. A. Hazzard, and J. L. Bohn, Phys. Rev. A {\bf 89}, 021601 (2014).
\bibitem{rancon2014} A. Ran\c{c}on and K. Levin, Phys. Rev. A {\bf 90}, 021602 (2014).

\bibitem{liu2015} X.-J. Liu, B. Mulkerin, L. He, and H. Hu, arXiv:1502.00345 (2015).

\bibitem{rem2013} B.S Rem, A.T Grier, I. Ferrier-Barbut, U. Eismann, T. Langen, N. Navon, L. Khaykovich, F. Werner, D.S. Petrov, F. Chevy, and C. Salomon, Phys. Rev. Lett. {\bf 110}, 163202 (2013).

\bibitem{lee1957} T. D. Lee, K. Huang, and C. N. Yang, Phys. Rev. {\bf 106}, 1135 (1957).

\bibitem{lee1957b}T. D. Lee and C. N. Yang, Phys. Rev. {\bf 105}, 1119 (1957).

\bibitem{Beliaev58} S. T. Beliaev, Sov. Phys. JETP. {\bf 7}, 289 (1958); Sov. Phys. JETP. {\bf 7}, 299 (1958).
\bibitem{wu1959} T. T. Wu, Phys. Rev. {\bf 115}, 1390 (1959).

\bibitem{sawada1959} K. Sawada, Phys. Rev. {\bf 116}, 1344 (1959).


\bibitem{Pines59}N. M. Hugenholtz and D. Pines, Phys. Rev. {\bf 116}, 489 (1959).

\bibitem{braaten2002} E. Braaten, H.-W. Hammer, and T. Mehen, Phys. Rev. Lett. {\bf 88}, 040401 (2002).

\bibitem{Fetter1971a} A. L. Fetter and J. D. Walecka, {\em Quantum Theory of Many-Particle Systems} (McGraw-Hill, San Francisco, 1971).





\bibitem{abrikosov1975methods} A. A. Abrikosov, L. P. Gorkov, and I. E. Dzyaloshinski, {\em Methods of Quantum Field Theory in Statistical Physics} (Courier Corporation, 1975).


\bibitem{cowell2002} S. Cowell, H. Heiselberg, I. E. Mazets, J. Morales, V. R. Pandharipande, and C. J. Pethick, Phys. Rev. Lett. {\bf 88}, 210403 (2002).

\bibitem{song2009} J. L. Song and F. Zhou, Phys. Rev. Lett. {\bf 103}, 025302 (2009).

\bibitem{diederix2011} J. M. Diederix, T. C. F. van Heijst, and H. T. C. Stoof, Phys. Rev. A {\bf 84}, 033618 (2011).


\bibitem{borzov2012} D. Borzov, M. S. Mashayekhi, S. Zhang, J.-L. Song, and F. Zhou, Phys. Rev. A {\bf 85}, 023620 (2012).
\bibitem{zhou2013} F. Zhou and M. S. Mashayekhi, Ann. Phys. {\bf 328}, 83 (2013).
\bibitem{pilati2005} S. Pilati, J. Boronat, J. Casulleras, and S. Giorgini, Phys. Rev. A {\bf 71}, 023605 (2005).
\bibitem{Mashayekhi2013} M. S. Mashayekhi, J.-S. Bernier, D. Borzov, J.-L. Song, and F. Zhou, Phys. Rev. Lett. {\bf 110}, 145301 (2013).

\bibitem{mahan2000many} G. D. Mahan, {\em Many-particle physics} (Springer Science \& Business Media, 2000).










\bibitem{stoof1994} H. T. C. Stoof, Phys. Rev. A {\bf 49}, 3824 (1994).

\bibitem{mueller2000} E. J. Mueller and G. Baym, Phys. Rev. A {\bf 62}, 053605 (2000).

\bibitem{jeon2002} G. S. Jeon, L. Yin, S. W. Rhee, and D. J. Thouless, Phys. Rev. A {\bf 66}, 011603 (2002).



\bibitem{li2012} W. Li and T.-L. Ho, Phys. Rev. Lett. {\bf 108}, 195301 (2012).


\bibitem{Dalfovo1999a} F. Dalfovo, S. Giorgini, L. P. Pitaevskii, and S. Stringari, Rev. Mod. Phys. {\bf 71}, 463 (1999).





\bibitem{fedichev1996} P.O. Fedichev, M.W. Reynolds, and G.V. Shlyapnikov, Phys. Rev. Lett. {\bf 77}, 2921 (1996).


\bibitem{nielsen1999} E. Nielsen and J.H. Macek, Phys. Rev. Lett. {\bf 83}, 1566 (1999).
\bibitem{esry1999} B.D. Esry, C.H. Greene, and J.P. Burke, Phys. Rev. Lett. {\bf 83}, 1751 (1999).
\bibitem{bedaque2000} P.F. Bedaque, E. Braaten, and H.-W. Hammer, Phys. Rev. Lett. {\bf 85}, 908 (2000).



\bibitem{Efimov70}V. Efimov, Phys. Lett. B. {\bf 33}, 563 (1970);








\bibitem{jiang2014} S.-J. Jiang, W.-M. Liu, G. W. Semenoff, and F. Zhou, Phys. Rev. A {\bf 89}, 033614 (2014).


\end{thebibliography}
\end{document}